\documentclass[aps,prl,reprint,amsmath,amssymb,groupedaddress]{revtex4-2}
\usepackage{graphicx}
\usepackage[colorlinks,
linkcolor=blue,
anchorcolor=blue,
citecolor=blue,urlcolor=blue]{hyperref}

\usepackage{xcolor}

\begin{document}

\preprint{APS/123-QED}

\title{Equilibrium Distributions for Strongly Nonlinear Many-Body Systems}

\author{Jialin Zhang}
\author{Yong Zhang}
\author{Hong Zhao}%
\email{zhaoh@xmu.edu.cn}
\affiliation{%
 \\Department of Physics, Xiamen University, Xiamen 361005, Fujian, China \\
}%
\date{\today}
\begin{abstract}
Obtaining equilibrium distributions of nonlinear systems is essential for accurately computing macroscopic observables. Conventional theoretical corrections are typically limited to weak nonlinearities, where interaction terms can be treated as effectively uncorrelated perturbations and the random phase approximation applies. In this Letter, we develop a framework to determine equilibrium distributions based on the generalized energy equipartition principle. Our approach recovers existing corrections in the weakly nonlinear regime and, crucially, remains valid for strong nonlinearities, where perturbative contributions become correlated and conventional approaches break down. Numerical simulations of the nonlinear Schrödinger equation, the Majda–McLaughlin–Tabak model, and the Fermi–Pasta–Ulam–Tsingou model demonstrate accurate corrections for nonlinearities more than an order of magnitude stronger than those accessible to conventional theories.   
\end{abstract}
\maketitle
The calculation of macroscopic observables from equilibrium distribution functions is a central task in statistical physics. In standard textbooks, these equilibrium distributions are typically derived under the harmonic approximation. However, nonlinear interactions are ubiquitous in real‑world systems. On the one hand, such interactions drive the system towards equilibrium \textsuperscript{\cite{onorato2015route,lvov2018double,wang2020wave,onorato2023wave,Wang2024Thermalization,Wang2024Thermalization2,lin2025universality}}, providing the theoretical foundation for the application of statistical physics. On the other hand, nonlinear interactions lead to deviations in the equilibrium distribution from the harmonic approximation. Perturbative corrections based on phonon Green's functions \textsuperscript{\cite{werthamer1970self,tripathi1974self,tadano2015self,tadano2022first,xiao2023anharmonic}} or tree-level Feynman diagrams \textsuperscript{\cite{boulware1968tree,driesse2024conservative,mougiakakos2024schwarzschild}} can, in principle, be used to renormalize the equilibrium distribution of nonlinear systems. However, in practice, these methods are confined to weak nonlinearity.

Other approaches for correcting stronger nonlinear effects have also been developed, including the variational approach\textsuperscript{\cite{liu2015renormalized,liu2016variational}}, the self-consistent harmonic approximation\textsuperscript{\cite{koehler1966theory,bruesch2012phonons}}, the self-consistent phonon theory\textsuperscript{\cite{he2008thermal,he2009origin,he2016quantum,masuki2022anharmonic}}, and trivial-resonance renormalization methods rooted in wave turbulence theory\textsuperscript{\cite{nazarenko2011wave,gershgorin2005renormalized,leisman2019effective,lee2009renormalized,gershgorin2007interactions,lvov2018double,chibbaro20184}}. These methods typically involve an appropriate transformation to re-decompose the Hamiltonian, introducing a new integrable Hamiltonian to replace the original harmonic one. The eigenfrequencies are then calculated using this new integrable Hamiltonian, replacing the frequency spectrum derived from the harmonic potential approximation. These methods are applicable when the residual interactions act as a weak, effectively uncorrelated perturbation with approximately Gaussian statistics, implying that the random phase approximation (RPA) holds and perturbative analytical correction formulas can be derived. However, in strongly nonlinear regimes where the residual interactions develop significant correlations and deviate from Gaussian statistics, the RPA breaks down, and a broadly applicable theoretical framework for computing equilibrium distribution functions is still lacking.

This Letter develops a general analytical framework for computing equilibrium distribution functions of nonlinear lattice systems. Unlike conventional approaches, our method is grounded in the generalized energy equipartition principle (GEEP)\textsuperscript{\cite{frisch1953equipartition,Pathria2016Statistical}}. For classical Hamiltonian systems in thermal equilibrium, this principle holds irrespective of the interaction strength. We therefore adopt the GEEP as a central constraint, upon which our analytical framework is constructed. In addition, we introduce an auxiliary generating function that streamlines the evaluation of the partition function. From the resulting equilibrium distributions, we further derive analytical corrections to the dispersion relations. Overall, the framework is applicable across weakly nonlinear regimes—where it is equivalent to existing corrections—and strongly nonlinear regimes, where conventional methods fail.

We apply this framework to the discrete nonlinear Schrödinger  (DNLS) equation\textsuperscript{\cite{kevrekidis2009discrete}}, the Majda–McLaughlin–Tabak (MMT) model\textsuperscript{\cite{majda1997one}}, and the Fermi–Pasta–Ulam–Tsingou-$\beta$ (FPUT-$\beta$) model. For each model, we derive the equilibrium distributions and the corresponding renormalized dispersion relations, and validate our predictions against numerical simulations.
The DNLS equation and MMT models have wide-ranging applications, including multimode nonlinear optical systems\textsuperscript{\cite{fibich2015nonlinear,zhong2023universality,ramos2023theory}}, waveguide lattices\textsuperscript{\cite{christodoulides2003discretizing}}, Bose–Einstein condensates\textsuperscript{\cite{vsindik2024sound,wachtler2016quantum}}, and water tank experiments\textsuperscript{\cite{tikan2022nonlinear}}. Their equilibrium states typically exhibit Rayleigh–Jeans (RJ) distributions\textsuperscript{\cite{connaughton2005condensation,sun2012observation,fusaro2019dramatic,pourbeyram2022direct,baudin2023observation,ramos2023theory}}. Existing approaches based on trivial-resonance renormalization\textsuperscript{\cite{nazarenko2011wave,chibbaro20184,leisman2022improved,lee2009renormalized,lee2013generation,lee2019reduced}} correct the eigenfrequencies, which in turn yield corrected RJ distributions\textsuperscript{\cite{tikan2022nonlinear}}.
In contrast, the FPUT-$\beta$ model is a canonical testbed for real-valued nonlinear lattices. A variety of schemes—including trivial-resonance renormalization\textsuperscript{\cite{gershgorin2005renormalized,lvov2018double}}, self-energy corrections for Green’s functions\textsuperscript{\cite{xu2008nonequilibrium}}, the self-consistent phonon method\textsuperscript{\cite{he2008thermal}}, and variational approaches\textsuperscript{\cite{liu2015renormalized}}—have been used to account for frequency renormalization and related nonlinear effects.

At thermal equilibrium, the equilibrium distribution of a classical Hamiltonian system satisfies the GEEP,
\begin{equation}
\left\langle x_i \,\frac{\partial (H-\mu N)}{\partial x_j} \right\rangle
= k_B T\, \delta_i^{\,j},
\label{GE}
\end{equation}
where $x_i$ denotes a canonical variable, $H$ is the Hamiltonian, $\mu$ is the chemical potential, $N$ is the particle number, $k_B$ is the Boltzmann constant, $T$ is the temperature, and $\delta_i^{\,j}$ is the Kronecker delta.

The Hamiltonian of the DNLS equation\textsuperscript{\cite{kevrekidis2009discrete}} is
\begin{equation}
H=\sum_{l\in\Lambda_L}\left(|\psi_{l+1}-\psi_l|^2+\frac{b}{2}|\psi_l|^4\right),
\label{HNLSE}
\end{equation}
where $\psi_l\in\mathbb{C}$ and $\Lambda_L=\{1,2,\ldots,L\}$. Assuming periodic boundary conditions, we expand $\psi_l$ in the eigenbasis of the linear part of Eq.~\eqref{HNLSE} via the discrete Fourier transform
$a_k=\frac{1}{\sqrt{L}}\sum_{l=1}^{L}\psi_l\,e^{-2\pi i k l/L},\qquad k=1,2,\ldots,L$.
This yields
\begin{equation}
H=\sum_{k}\omega_k^{(0)}a_k a_k^*+\lambda\sum_{1234} a_1 a_2 a_3^{*} a_4^{*}\,\delta_{34}^{12},
\label{Hk}
\end{equation}
where $\omega_k^{(0)}=4\sin^2\!\left(\frac{k\pi}{L}\right)$ is the linear dispersion relation, $\lambda=b/(2L)$, and $\sum_{1234}$ is shorthand for the sum over $(k_1,k_2,k_3,k_4)$. The Kronecker symbol $\delta_{34}^{12}$ enforces the resonance condition,  $\delta^{12}_{34}=1$, for $ (k_1 + k_2 - k_3 - k_4) \mod L = 0$, and $\delta^{12}_{34}=0$ otherwise.

This system possesses two conserved quantities: the Hamiltonian $H$ and the total particle number
$N=\sum_{l}|\psi_l|^2$. We therefore work in the grand-canonical ensemble, with partition function $Z=\int \exp\!\big[-\beta\,(H-\mu N)\big]\,d\Omega$, where $\beta=\frac{1}{k_B T}$.  Treating $(a_k, a_k^*)$ as generalized coordinates and considering only the linear part of the Hamiltonian, the GEEP yields:
\begin{equation}
n_k^{(0)}=\frac{k_B T}{\omega_k^{(0)}-\mu},
\label{RJwr}
\end{equation}
where $n_k=\langle a_k a_k^*\rangle$. This is the equilibrium distribution that a system conserving energy and particle number should exhibit within the harmonic approximation.

\begin{figure}[t]
    \centering
    \includegraphics[scale=0.55]{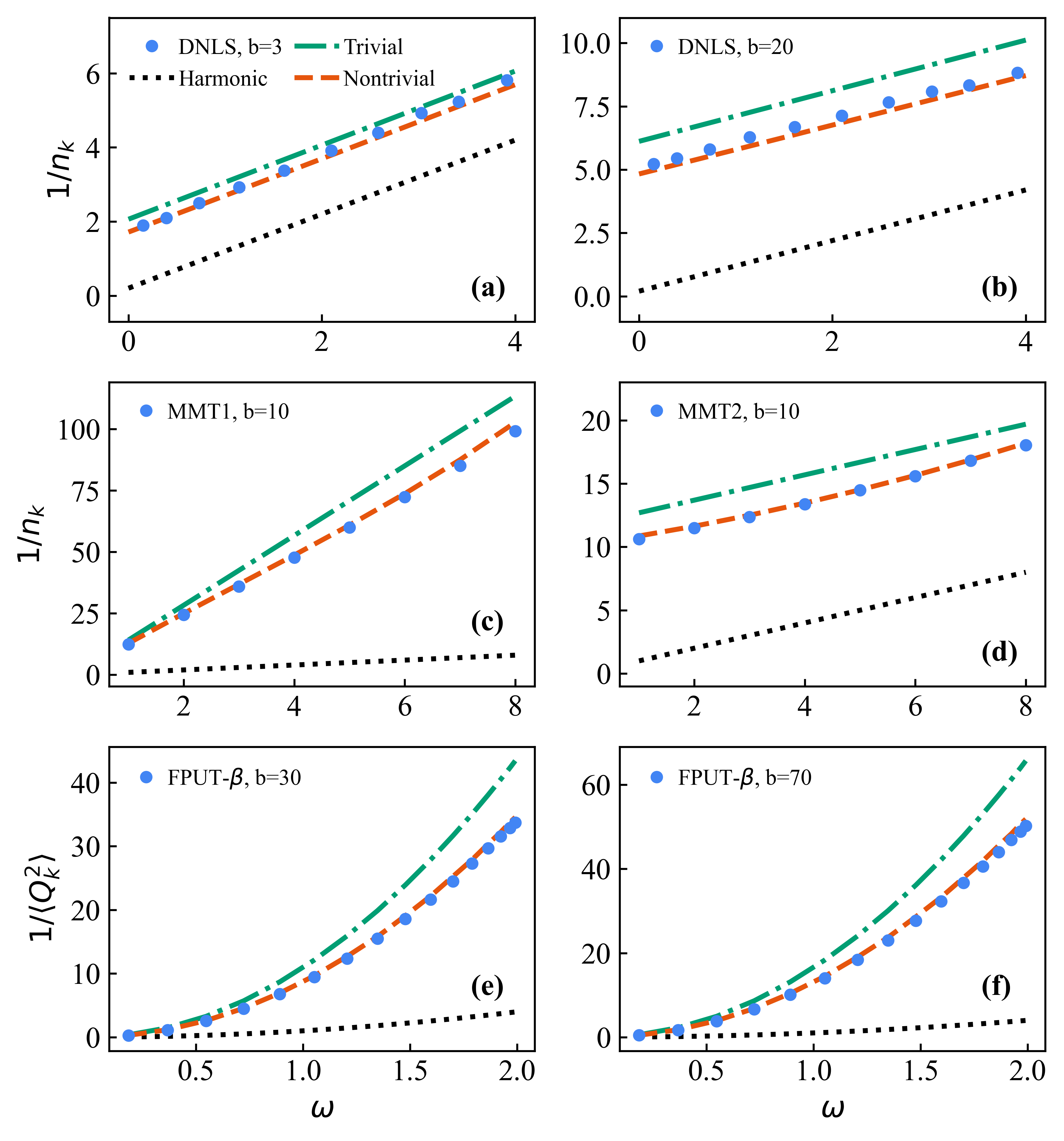}
    \caption{
Comparison between numerical results (solid circles) and the harmonic-approaxmation prediction (black dotted line), the trivial-resonance prediction (green dash-dotted line), and our full theoretical prediction including nontrivial-resonance corrections (red dashed line). Panels (a,b): DNLS equation with $b=3$ and $b=20$; (c): MMT1 with $b=10$; (d): MMT2 with $b=10$; and (e,f): FPUT-$\beta$ with $b=30$ and $b=70$.}
    \label{NLSERJFIG}
\end{figure}

In Figs.~\ref{NLSERJFIG} (a) and \ref{NLSERJFIG} (b), we show numerically obtained equilibrium distributions for different nonlinearity strengths $b$ (with $T=1$, $\mu=-0.2$, $L=256$, and $k_B=1$). We observe significant deviations from the prediction of Eq.~\eqref{RJwr}. Details of the simulation algorithm, as well as the theoretical derivations used throughout this Letter, are provided in the Supplemental Material (SM)\textsuperscript{\cite{SuppMat}}.

When nonlinear terms are included, four-wave resonances (quartets) can be divided into trivial and non-trivial resonances \textsuperscript{\cite{nazarenko2011wave}}. The former occurs without any momentum exchange, involving $(k_1 = k_3, k_2 = k_4)$ and $(k_1 = k_4, k_2 = k_3)$. The latter occurs with momentum exchange processes, including Umklapp processes. These resonances take the form $k_1+k_2\equiv k_3+k_4\ (\mathrm{mod}\ L)$. Applying the GEEP, we obtain:
    \begin{equation}
        \begin{aligned}
        n_k (\omega^{(1)}_k-\mu) +b/L\sum^{**}_{123} \langle a_1 a_2^* a_3^* a_k\rangle\delta_{12}^{3k}=k_BT,
        \end{aligned}
        \label{RJ}
    \end{equation}
where the term $\omega^{(1)}_k = \omega^{(0)}_k + \frac{2bN}{L}$ represents the contribution from trivial resonances, and $\sum^{**}$ denotes a summation excluding the trivial resonances. The summation term in the equation thus corresponds to the contribution from non-trivial resonances. If we assume that the non-trivial resonances satisfy RPA, the summation term will be zero, leading to the distribution:
    \begin{equation}
    n^{(1)}_k=\frac{k_B T}{\omega^{(1)}_k-\mu}.
    \label{RJtr}
    \end{equation}
It can be observed that this form is consistent with previous results\textsuperscript{\cite{nazarenko2011wave,lee2009renormalized}}. 

However, the assumption that non-trivial resonances can be neglected and the validity of Eq.~\ref{RJtr} rely on Wick-type factorization implied by RPA. Wick’s selection rule gives: $\langle a_1 a_2 a_3^* a_4^*\rangle = n_1 n_2 (\delta_{1}^{3}\delta_{2}^{4}+\delta_{1}^{4}\delta_{2}^{3}) + (\langle |a_1|^4 \rangle -2 n_1^2) (\delta_{1}^{2}\delta_{1}^{3}\delta_{1}^{4})$\textsuperscript{\cite{nazarenko2011wave}}. 
This leads to $\sum_{123}\langle a_1 a_2 a_3^{} a_k^{}\rangle\delta^{12}_{3k}=2Nn_k$ for trivial resonances, provided the system is sufficiently large so that the last term ($\langle |a_1|^4 \rangle -2 n_1^2) (\delta_{1}^{2}\delta_{1}^{3}\delta_{1}^{4})$) is negligible, whereas $\sum_{1234}\langle {a_1 a_2 a^*_3 a^*_4 }\rangle\delta^{12}_{34}=0$ for non‑trivial resonances. One then recovers Eq.~\ref{RJtr}.

As shown in Fig.~\ref{NLSERJFIG}(a) and \ref{NLSERJFIG}(b), the trivial-resonance correction provides a partial improvement of the theoretical predictions. However, the agreement with numerics is still far from satisfactory. Further numerical tests indicate that this correction remains accurate only for $b\lesssim 1$.

\begin{figure}[t]
\centering
\includegraphics[scale=0.5]{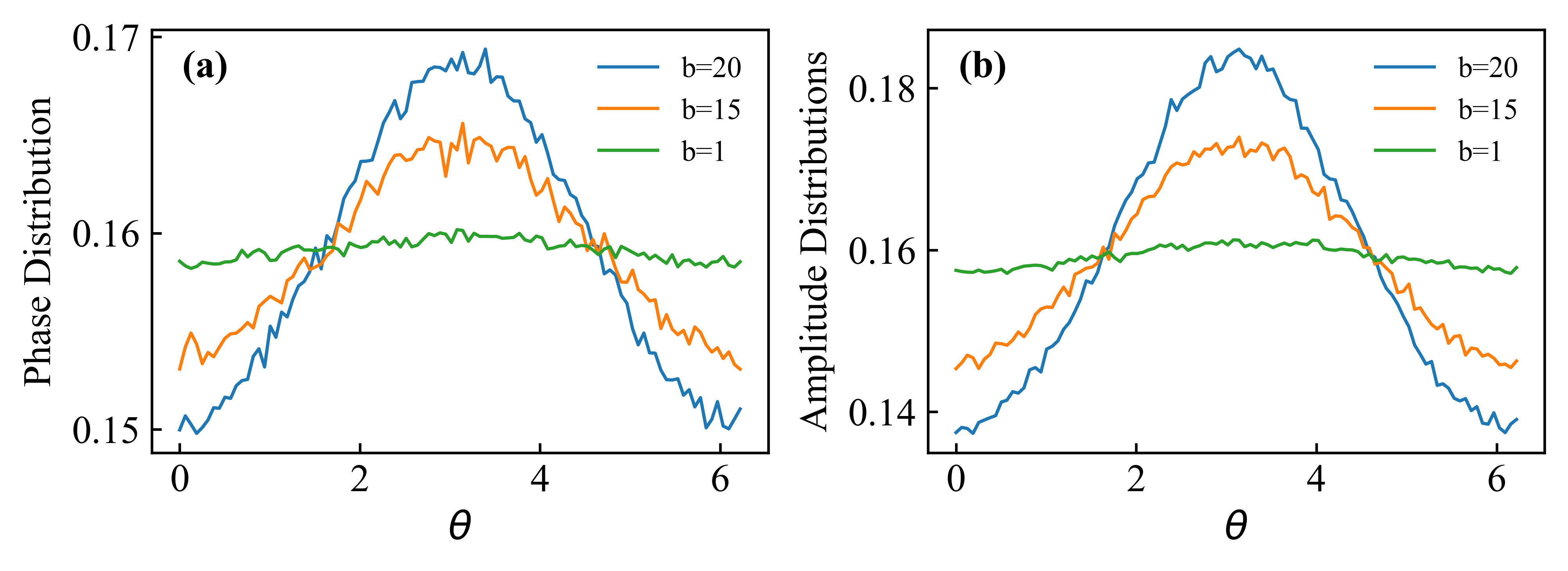}
\caption{Phase distribution (a) and amplitude distribution (b) for a representative nontrivial resonant quartet $(k_1,k_2,k_3,k_4)=(3,4,2,5)$ at different nonlinearity strengths. The curves at $\theta=\pi$ correspond, from top to bottom, to $b=20$, $b=15$, and $b=1$, respectively.
}
\label{4thReIm}
\end{figure}

These discrepancies indicate that the RPA and Wick-type factorization no longer hold in the strongly nonlinear regime. We consider a resonant quartet of modes $(k_1,k_2,k_3,k_4)=(3,4,2,5)$, which constitutes a nontrivial resonance because the  condition $k_1+k_2=k_3+k_4$ is satisfied. Writing $a_j=|a_j|e^{i\zeta_j}$, the phase of the product $a_1 a_2 a_3^{*} a_4^{*}$ is $\theta=\zeta_1+\zeta_2-\zeta_3-\zeta_4$. In Fig.~\ref{4thReIm}(a), we plot the sampled distribution of $\theta$ over $[0,2\pi)$. If the RPA holds, $\theta$ should be uniformly distributed on this interval; instead, we observe a clear deviation from uniformity. As the nonlinearity decreases, the distribution gradually approaches the RPA prediction. Likewise, under the RPA the amplitude $|a_1 a_2 a_3^{*} a_4^{*}|$ should be statistically independent of $\theta$, whereas Fig.~\ref{4thReIm}(b) shows a pronounced amplitude--phase dependence.

To incorporate corrections induced by nontrivial resonances, we reorganize the Hamiltonian by absorbing the trivial-resonance frequency shift into the quadratic part, which yields
\begin{equation}
    \begin{aligned}
        H
        = \sum_k \omega^{(1)}_k\, a_k a_k^{*}
        + \lambda \sum_{1234}^{**} a_1 a_2 a_3^{*} a_4^{*}\, \delta^{12}_{34}.
    \end{aligned}
    \label{HDNLSk}
\end{equation}
In this representation, the remaining nonintegrable contribution to the Hamiltonian arises solely from the interaction terms associated with nontrivial resonances.

We then introduce two sets of real external sources, $\{J_k^{1},J_k^{2}\}\subset\mathbb{R}$, and define a source-dependent partition function (generating functional):
\begin{equation}
    \begin{aligned}
        Z(J)
        = \int \exp\!\left[-\beta\,(H-\mu N) + \sum_k \bigl(J^1_k a_k + J^2_k a_k^{*}\bigr)\right]\, d\Omega ,
    \end{aligned}
    \label{PartitionFunction}
\end{equation}
where $d\Omega=\prod_k da_k$. Setting $J^1_k=J^2_k=0$ recovers the partition function $Z(0)$ of the Hamiltonian.
From this generating functional, we obtain
\begin{equation}
    \begin{aligned}
        \langle a_1 a_2 a_3^* a_4^* \rangle
        = \frac{1}{Z(0)}
        \frac{\partial^4 Z(J)}{\partial J^1_1\, \partial J^1_2\, \partial J^2_3\, \partial J^2_4}\Bigg|_{J^1_k=J^2_k=0},
    \end{aligned}
\end{equation}.

Expanding the nontrivial resonance part of the Hamiltonian in Eq.~\eqref{PartitionFunction} yields
\begin{equation}
    \begin{aligned}
        Z(J)
        =& \prod_{k}\frac{\pi}{\beta(\omega^{(1)}_k-\mu)}
        \exp\!\left(\frac{J^1_k J^2_k}{\beta(\omega^{(1)}_k-\mu)}\right) \\
       & - \beta\lambda \sum_{1234}^{**}\delta^{12}_{34}\,
        \frac{\partial^4 Z^{(1)}(J)}{\partial J_1^{2}\, \partial J_2^{2}\, \partial J_3^{1}\, \partial J_4^{1}}
        + \cdots.
    \end{aligned}
    \label{PartitionResult}
\end{equation}
Setting $J^1_k=J^2_k=0$, the first term gives the trivial-resonance contribution $Z^{(1)}(0)= \pi\prod_k n_k^{(1)}$. 
For a nontrivial resonance quartet, one has $\langle a_1 a_2 a_3^* a_4^* \rangle^{(1)}=0$, where the superscript $(1)$ denotes the ensemble average taken with respect to $Z^{(1)}(0)$. 

The second term accounts for contributions from nontrivial interactions. Taking the ensemble average with respect to the second term, we then obtain the high-order correction:
\begin{equation}
\begin{aligned}
    \langle a_1 a_2 a_3^* a_4^* \rangle^{(2)}
    &\simeq -4\beta\lambda\,\delta^{12}_{34}\,
    \prod_{k\in\{1,2,3,4\}}\frac{1}{\beta(\omega^{(1)}_k-\mu)}.
\end{aligned}
\label{4moment}
\end{equation}
This correction accounts for the leading contribution of nontrivial resonances, with the higher-order terms in Eq.~\eqref{PartitionResult} neglected (see SM\textsuperscript{\cite{SuppMat}}). Applying the GEEP then yields
\begin{equation}
    \begin{aligned}
        n_k^{(2)}
        = \frac{\beta^{-1}
        + 8\beta \lambda^2 \sum_{123}^{**} n_1^{(1)} n_2^{(1)} n_3^{(1)} n_k^{(1)}\, \delta^{12}_{3k}}
        {\omega^{(1)}_k-\mu},
    \end{aligned}
    \label{RJNLSE}
\end{equation}
which explicitly incorporates the nontrivial-resonance correction. All quantities on the right-hand side are analytically computable.

Figures~\ref{NLSERJFIG}(a) and \ref{NLSERJFIG}(b) show that this correction agrees quantitatively with numerical results for at least $b \gtrsim 20$, extending the range of nonlinear strengths amenable to analytical correction by more than an order of magnitude beyond the trivial-resonance approach. Numerical tests further confirm that this correction remains accurate for stronger nonlinear interactions and across a wider parameter range (see SM\textsuperscript{\cite{SuppMat}}).
    
    \begin{figure}[]
        \centering
        \includegraphics[scale=0.55]{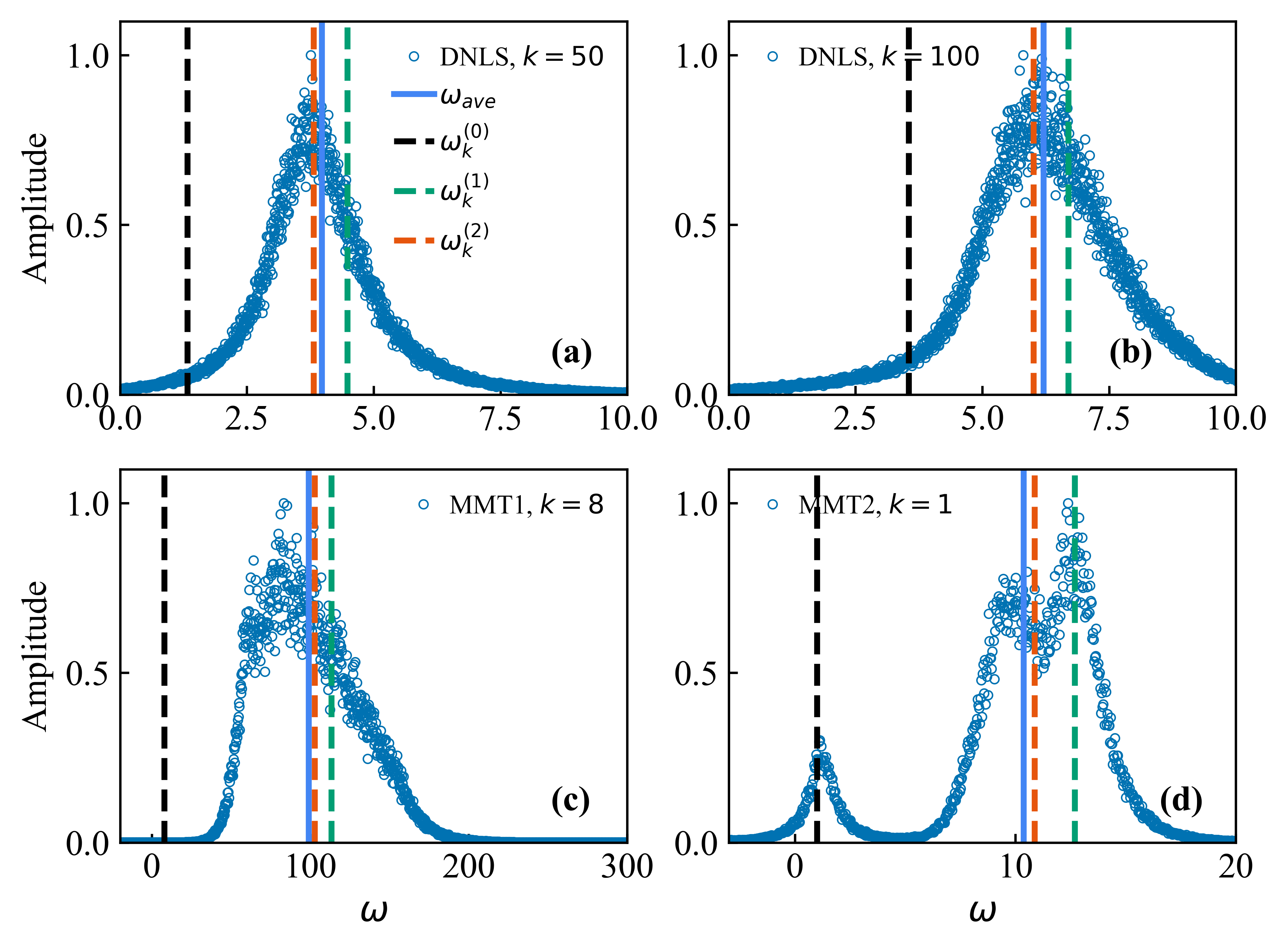}
        \caption{For the DNLS equation at $b = 7$ for $k = 50$ (a) and $k = 100$ (b); MMT1 model at $b = 10$, $k = 8$ (c); MMT2 model at $b = 10$, $k = 1$ (d): Frequency spectrum (dots), average frequency (blue solid line), harmonic approximation prediction frequency (black dashed line), trivial-resonance frequency prediction (green dashed line), and our frequency prediction (red dashed line). Spectra are normalized by their peak values.}
        \label{frequency}
    \end{figure}

We then turn to the MMT-like models with the Hamiltonian
  \begin{equation}
        \begin{aligned}
            H=&\sum_k \omega_k^{(0)} a_k a^*_k + a \sum_{123} a_1 a_2 a^*_3 \delta^{12}_{3}  + a \sum_{123} a_1 a^*_2 a^*_3\delta^{1}_{23} \\
            & + \frac{b}{2} \sum_{1234}   |k_1 k_2 k_3 k_4|^{c/4} a_1 a_2 a^*_3 a^*_4 \delta^{12}_{34} .\\
        \end{aligned}
    \end{equation}
Here, $\omega_k^{(0)} = |k|^d$ is the dispersion relation. The cubic terms represent three-wave processes satisfying $k_1+k_2= k_3$ and $k_1=k_2+ k_3$. The quartic term includes all $2-2$ four-wave processes. The parameters $a$ and $b$ control the interaction strengths of the cubic and quartic terms, respectively, while $c$ and $d>0$ are real-valued constants.

Within the harmonic approximation, the GEEP yields the distribution that satisfies:
\begin{equation}
    n_k^{(0)} = \frac{k_B T}{\omega^{(0)}_k}.
    \label{R1}
\end{equation}
There are no trivial resonances in the cubic terms. Moreover, if we also neglect the nontrivial resonances in the quartic term, the GEEP gives the trivial-resonance correction:
    \begin{equation}
        \begin{aligned}
 n_k^{(1)}  = \frac{k_B T}{\omega^{(1)}_k}
        \end{aligned}
        \label{R1}
    \end{equation}
with $\omega^{(1)}_k = |k|^d + \left( 2 b \sum_{k'} |k'|^{c/2} \langle |a_{k'}|^2 \rangle \right) |k|^{c/2}$. By reorganizing the Hamiltonian, absorbing the trivial-resonance frequency shift into the quadratic part, we obtain the equilibrium distribution that includes the nontrivial-resonance correction:
\begin{equation}
        \begin{aligned}
         n^{(2)}_k \simeq  \frac{k_B T + \beta n_k^{(1)} F_{123k} }{\omega^{(1)}_k},
        \end{aligned}
        \label{RJMMT34}
    \end{equation}
where $F_{123k} = \sum_{12} 2 a^2 n_1^{(1)} n_2^{(1)} \delta^{12}_k + \sum_{12} 4 a^2 n_1^{(1)} n_2^{(1)} \delta^{1}_{2k} + \sum_{123}^{**} 2 b^2 |k_1 k_2 k_3 k_4|^{c/2} n_1^{(1)} n_2^{(1)} n_3^{(1)} \delta^{12}_{3k}$. 

The numerical simulations again indicate that the nontrivial-resonance correction yields a highly accurate equilibrium distribution, significantly outperforming the trivial-resonance correction in the strong-nonlinearity regime. Fig.~\ref{NLSERJFIG}(c) shows example for the model ($a=0$) with $b=10$ (shorten as MMT1 model). The other parameters are $c=2$ and $d=1$. Similarly, Fig.~\ref{NLSERJFIG}(d) shows example for the model ($a=\sqrt{b}$) with $b=10$  (shorten as MMT2 model). The other parameters are $c=0$ and $d=1$. The first $8$ norm modes ($k=1,2,...,8$) are involved in the simulations. Numerical tests for even larger $k$ lead to the same conclusion (see SM\textsuperscript{\cite{SuppMat}}).

Once the equilibrium distribution is obtained, we can derive the average frequency of the $k$th mode, defined as
\begin{equation}
\bar{\omega}_{k}
= -\,\frac{\int \omega\, S_k(\omega)\, d\omega}{\int S_k(\omega)\, d\omega},
\end{equation}
where $S_k(\omega)$ denotes the power spectral density of $a_k(t)$. This expression yields
\begin{equation}
\begin{aligned}
{\omega}_{k}^{(2)} = \frac{k_B T}{n_k^{(2)}} + \mu,
\end{aligned}
\label{omegaave}
\end{equation}
when nontrivial resonances are considered. Here, $\mu \neq 0$ for DNLS equation, while $\mu = 0$ for the MMT models. Replacing $n_k^{(2)}$ with $n_k^{(1)}$ and $n_k^{(0)}$, we obtain the frequency corrections for the trivial resonance, ${\omega}_{k}^{(1)}$, and the harmonic approximation, ${\omega}_{k}^{(0)}$, respectively.

Figure~\ref{frequency} compares the frequency predictions from the harmonic approximation, the trivial-resonance prediction, and the nontrivial-resonance prediction. The corresponding frequency distributions from numerical simulations are also shown. We find that the nontrivial-resonance correction agrees closely with the numerically computed average frequencies, even at large nonlinearities, whereas both the harmonic approximation and the trivial-resonance prediction exhibit substantial deviations. In particular, the trivial-resonance prediction systematically overestimates the frequency shift, and the nontrivial-resonance contribution compensates for this overestimate.

These plots demonstrate that, although the frequency distributions no longer retain a standard Lorentzian profile---and Fig.~\ref{frequency}(d) even shows that the frequency spectrum of a mode can split into three peaks when the cubic nonlinearity is present---the prediction for the mean frequency remains nearly quantitative. Fig.~\ref{frequency}(d) revels an additional intriguing feature: the three peaks associated with the $k$th normal mode are located close to the harmonic approximation predicted eigenfrequency, the trivial-resonance-corrected frequency, and the mean frequency predicted with the nontrivial-resonance correction, respectively.

Finally, we report the results for the FPUT-$\beta$ model. Its Hamiltonian is given by:
    \begin{equation}
        \begin{aligned}
            H = \sum_j \frac{p_j^2}{2} + \frac{(q_{j+1} - q_j)^2}{2} +  \frac{b(q_{j+1} - q_j)^4}{4}.
        \end{aligned}
    \end{equation}
Using the same procedure as for the previous two models and applying the GEEP, we then derive the correction due to nontrivial resonances. Fig.~\ref{NLSERJFIG}(e) and ~\ref{NLSERJFIG}(f) shows the results under fixed boundary conditions, where the theoretical predictions remain consistent with numerical simulations across the entire range from weak to strong nonlinearity. Here we show numerical simulations for $T=1$ with $j=16$ particles. Numerical tests with $j=64$ particles lead to the same conclusion (see SM\textsuperscript{\cite{SuppMat}}). Theoretical analysis for this model is more involved, as it requires accounting for multiple types of nontrivial resonances. 

In summary, in the very strong nonlinear regime, nontrivial resonances can no longer be treated as weakly correlated interactions; consequently, RPA breaks down. In this regime, the nonlinear effects of these resonances must be explicitly accounted for. Building on an analytical framework grounded in the GEEP, we recover equilibrium distribution corrections for the harmonic approximation and trivial resonances, and derive equilibrium distributions and dispersion relations that remain valid even for very strong nonlinear interactions. This is achieved by introducing a generating functional for the partition function and renormalizing the Hamiltonian, absorbing the trivial-resonance frequency shift into the integrable part. Numerical simulations of several representative models demonstrate the broad applicability of our approach. Given the widespread presence of strong nonlinearity in applied systems, our analytical framework has significant potential for broader use.

% \nocite{*} 

% \bibliographystyle{unsrt} 
\bibliography{referenceNEW}% Produces the bibliography via BibTeX.

@PREAMBLE{
 "\providecommand{\noopsort}[1]{}" 
 # "\providecommand{\singleletter}[1]{#1}%" 
}

@article{lvov2018double,
  title = {Double Scaling in the Relaxation Time in the $\ensuremath{\beta}$-Fermi-Pasta-Ulam-Tsingou Model},
  author = {Lvov, Yuri V. and Onorato, Miguel},
  journal = {Phys. Rev. Lett.},
  volume = {120},
  issue = {14},
  pages = {144301},
  numpages = {5},
  year = {2018},
  month = {Apr},
  publisher = {American Physical Society},
  doi = {10.1103/PhysRevLett.120.144301},
  url = {https://link.aps.org/doi/10.1103/PhysRevLett.120.144301}
}

@article{boulware1968tree,
  title = {Tree Graphs and Classical Fields},
  author = {Boulware, David G. and Brown, Lowell S.},
  journal = {Phys. Rev.},
  volume = {172},
  issue = {5},
  pages = {1628--1631},
  numpages = {0},
  year = {1968},
  month = {Aug},
  publisher = {American Physical Society},
  doi = {10.1103/PhysRev.172.1628},
  url = {https://link.aps.org/doi/10.1103/PhysRev.172.1628}
}

@article{driesse2024conservative,
  title = {Conservative Black Hole Scattering at Fifth Post-Minkowskian and First Self-Force Order},
  author = {Driesse, Mathias and Jakobsen, Gustav Uhre and Mogull, Gustav and Plefka, Jan and Sauer, Benjamin and Usovitsch, Johann},
  journal = {Phys. Rev. Lett.},
  volume = {132},
  issue = {24},
  pages = {241402},
  numpages = {9},
  year = {2024},
  month = {Jun},
  publisher = {American Physical Society},
  doi = {10.1103/PhysRevLett.132.241402},
  url = {https://link.aps.org/doi/10.1103/PhysRevLett.132.241402}
}

@article{mougiakakos2024schwarzschild,
  title = {Schwarzschild Metric from Scattering Amplitudes to All Orders in ${G}_{N}$},
  author = {Mougiakakos, Stavros and Vanhove, Pierre},
  journal = {Phys. Rev. Lett.},
  volume = {133},
  issue = {11},
  pages = {111601},
  numpages = {9},
  year = {2024},
  month = {Sep},
  publisher = {American Physical Society},
  doi = {10.1103/PhysRevLett.133.111601},
  url = {https://link.aps.org/doi/10.1103/PhysRevLett.133.111601}
}

@article{wang2024thermalization,
title = {Thermalization of one-dimensional classical lattices: beyond the weakly interacting regime},
doi = {10.1088/1572-9494/ad696d},
url = {https://doi.org/10.1088/1572-9494/ad696d},
year = {2024},
month = {sep},
publisher = {IOP Publishing},
volume = {76},
number = {11},
pages = {115601},
author = {Wang, Zhen and Fu, Weicheng and Zhang, Yong and Zhao, Hong},
journal = {Commun. Theor. Phys.},
}

@article{xiao2023anharmonic,
  title = {Anharmonic phonon behavior via irreducible derivatives: Self-consistent perturbation theory and molecular dynamics},
  author = {Xiao, Enda and Marianetti, Chris A.},
  journal = {Phys. Rev. B},
  volume = {107},
  issue = {9},
  pages = {094303},
  numpages = {9},
  year = {2023},
  month = {Mar},
  publisher = {American Physical Society},
  doi = {10.1103/PhysRevB.107.094303},
  url = {https://link.aps.org/doi/10.1103/PhysRevB.107.094303}
}

@article{werthamer1970self,
  title = {Self-Consistent Phonon Formulation of Anharmonic Lattice Dynamics},
  author = {Werthamer, N. R.},
  journal = {Phys. Rev. B},
  volume = {1},
  issue = {2},
  pages = {572--581},
  numpages = {0},
  year = {1970},
  month = {Jan},
  publisher = {American Physical Society},
  doi = {10.1103/PhysRevB.1.572},
  url = {https://link.aps.org/doi/10.1103/PhysRevB.1.572}
}

@article{tadano2015self,
  title = {Self-consistent phonon calculations of lattice dynamical properties in cubic ${\mathrm{SrTiO}}_{3}$ with first-principles anharmonic force constants},
  author = {Tadano, Terumasa and Tsuneyuki, Shinji},
  journal = {Phys. Rev. B},
  volume = {92},
  issue = {5},
  pages = {054301},
  numpages = {10},
  year = {2015},
  month = {Aug},
  publisher = {American Physical Society},
  doi = {10.1103/PhysRevB.92.054301},
  url = {https://link.aps.org/doi/10.1103/PhysRevB.92.054301}
}

@article{tripathi1974self,
	title = {Self-energy of phonons in an anharmonic crystal to {O}(δ4)},
	volume = {21},
	issn = {1826-9877},
	url = {https://doi.org/10.1007/BF02737485},
	doi = {10.1007/BF02737485},
	number = {2},
	journal = {Il Nuovo Cimento B},
	author = {Tripathi, R. S. and Pathak, K. N.},
	month = {jun},
	year = {1974},
	pages = {289--302},
}

@article{tadano2022first,
  title = {First-Principles Phonon Quasiparticle Theory Applied to a Strongly Anharmonic Halide Perovskite},
  author = {Tadano, Terumasa and Saidi, Wissam A.},
  journal = {Phys. Rev. Lett.},
  volume = {129},
  issue = {18},
  pages = {185901},
  numpages = {7},
  year = {2022},
  month = {Oct},
  publisher = {American Physical Society},
  doi = {10.1103/PhysRevLett.129.185901},
  url = {https://link.aps.org/doi/10.1103/PhysRevLett.129.185901}
}

@article{masuki2022anharmonic,
  title = {Anharmonic Gr\"uneisen theory based on self-consistent phonon theory: Impact of phonon-phonon interactions neglected in the quasiharmonic theory},
  author = {Masuki, Ryota and Nomoto, Takuya and Arita, Ryotaro and Tadano, Terumasa},
  journal = {Phys. Rev. B},
  volume = {105},
  issue = {6},
  pages = {064112},
  numpages = {17},
  year = {2022},
  month = {Feb},
  publisher = {American Physical Society},
  doi = {10.1103/PhysRevB.105.064112},
  url = {https://link.aps.org/doi/10.1103/PhysRevB.105.064112}
}

@article{he2016quantum,
  title = {Quantum thermal transport through anharmonic systems: A self-consistent approach},
  author = {He, Dahai and Thingna, Juzar and Wang, Jian-Sheng and Li, Baowen},
  journal = {Phys. Rev. B},
  volume = {94},
  issue = {15},
  pages = {155411},
  numpages = {6},
  year = {2016},
  month = {Oct},
  publisher = {American Physical Society},
  doi = {10.1103/PhysRevB.94.155411},
  url = {https://link.aps.org/doi/10.1103/PhysRevB.94.155411}
}

@book{bruesch2012phonons,
  title={Phonons: Theory and experiments I},
  author={Br{\"u}esch, Peter},
  year={1982},
  publisher={Springer Berlin, Heidelberg},
}

@article{koehler1966theory,
  title = {Theory of the Self-Consistent Harmonic Approximation with Application to Solid Neon},
  author = {Koehler, Thomas R.},
  journal = {Phys. Rev. Lett.},
  volume = {17},
  issue = {2},
  pages = {89--91},
  numpages = {0},
  year = {1966},
  month = {Jul},
  publisher = {American Physical Society},
  doi = {10.1103/PhysRevLett.17.89},
  url = {https://link.aps.org/doi/10.1103/PhysRevLett.17.89}
}

@article{he2008thermal,
  title = {Thermal conductivity of anharmonic lattices: Effective phonons and quantum corrections},
  author = {He, Dahai and Buyukdagli, Sahin and Hu, Bambi},
  journal = {Phys. Rev. E},
  volume = {78},
  issue = {6},
  pages = {061103},
  numpages = {8},
  year = {2008},
  month = {Dec},
  publisher = {American Physical Society},
  doi = {10.1103/PhysRevE.78.061103},
  url = {https://link.aps.org/doi/10.1103/PhysRevE.78.061103}
}

@article{he2009origin,
  title = {Origin of negative differential thermal resistance in a chain of two weakly coupled nonlinear lattices},
  author = {He, Dahai and Buyukdagli, Sahin and Hu, Bambi},
  journal = {Phys. Rev. B},
  volume = {80},
  issue = {10},
  pages = {104302},
  numpages = {6},
  year = {2009},
  month = {Sep},
  publisher = {American Physical Society},
  doi = {10.1103/PhysRevB.80.104302},
  url = {https://link.aps.org/doi/10.1103/PhysRevB.80.104302}
}

@article{gershgorin2005renormalized,
  title = {Renormalized Waves and Discrete Breathers in $\ensuremath{\beta}$-Fermi-Pasta-Ulam Chains},
  author = {Gershgorin, Boris and Lvov, Yuri V. and Cai, David},
  journal = {Phys. Rev. Lett.},
  volume = {95},
  issue = {26},
  pages = {264302},
  numpages = {4},
  year = {2005},
  month = {Dec},
  publisher = {American Physical Society},
  doi = {10.1103/PhysRevLett.95.264302},
  url = {https://link.aps.org/doi/10.1103/PhysRevLett.95.264302}
}

@article{leisman2019effective,
  title = {Effective dispersion in the focusing nonlinear Schr\"odinger equation},
  author = {Leisman, Katelyn Plaisier and Zhou, Douglas and Banks, J. W. and Kova\ifmmode \check{c}\else \v{c}\fi{}i\ifmmode \check{c}\else \v{c}\fi{}, Gregor and Cai, David},
  journal = {Phys. Rev. E},
  volume = {100},
  issue = {2},
  pages = {022215},
  numpages = {19},
  year = {2019},
  month = {Aug},
  publisher = {American Physical Society},
  doi = {10.1103/PhysRevE.100.022215},
  url = {https://link.aps.org/doi/10.1103/PhysRevE.100.022215}
}

@article{lee2009renormalized,
  title = {Renormalized Resonance Quartets in Dispersive Wave Turbulence},
  author = {Lee, Wonjung and Kova\ifmmode \check{c}\else \v{c}\fi{}i\ifmmode \check{c}\else \v{c}\fi{}, Gregor and Cai, David},
  journal = {Phys. Rev. Lett.},
  volume = {103},
  issue = {2},
  pages = {024502},
  numpages = {4},
  year = {2009},
  month = {Jul},
  publisher = {American Physical Society},
  doi = {10.1103/PhysRevLett.103.024502},
  url = {https://link.aps.org/doi/10.1103/PhysRevLett.103.024502}
}

@article{gershgorin2007interactions,
title = {Interactions of renormalized waves in thermalized Fermi-Pasta-Ulam chains},
author = {Gershgorin, Boris and Lvov, Yuri V. and Cai, David},
journal = {Phys. Rev. E},
volume = {75},
issue = {4},
pages = {046603},
numpages = {15},
year = {2007},
month = {Apr},
publisher = {American Physical Society},
doi = {10.1103/PhysRevE.75.046603},
url = {https://link.aps.org/doi/10.1103/PhysRevE.75.046603}
}

@book{nazarenko2011wave,
  title={Wave turbulence},
  author={Nazarenko, Sergey},
  year={2011},
  publisher={Springer Berlin, Heidelberg}
}

@article{chibbaro20184,
  title = {4-wave dynamics in kinetic wave turbulence},
  journal = {Physica D: Nonlinear Phenomena},
  volume = {362},
  pages = {24-59},
  year = {2018},
  issn = {0167-2789},
  doi = {https://doi.org/10.1016/j.physd.2017.09.001},
  url = {https://www.sciencedirect.com/science/article/pii/S0167278917301847},
  author = {Sergio Chibbaro and Giovanni Dematteis and Lamberto Rondoni}
}

@article{lin2025universality,
  title = {Universality classes of thermalization and energy diffusion},
  author = {Lin, Wei and Fu, Weicheng and Wang, Zhen and Zhang, Yong and Zhao, Hong},
  journal = {Phys. Rev. E},
  volume = {111},
  issue = {2},
  pages = {024122},
  numpages = {6},
  year = {2025},
  month = {Feb},
  publisher = {American Physical Society},
  doi = {10.1103/PhysRevE.111.024122},
  url = {https://link.aps.org/doi/10.1103/PhysRevE.111.024122}
}

@article{wang2020wave,
  title = {Wave-Turbulence Origin of the Instability of Anderson Localization against Many-Body Interactions},
  author = {Wang, Zhen and Fu, Weicheng and Zhang, Yong and Zhao, Hong},
  journal = {Phys. Rev. Lett.},
  volume = {124},
  issue = {18},
  pages = {186401},
  numpages = {6},
  year = {2020},
  month = {May},
  publisher = {American Physical Society},
  doi = {10.1103/PhysRevLett.124.186401},
  url = {https://link.aps.org/doi/10.1103/PhysRevLett.124.186401}
}

@article{onorato2023wave,
  title = {Wave Turbulence and thermalization in one-dimensional chains},
  journal = {Phys. Rep.},
  volume = {1040},
  pages = {1-36},
  year = {2023},
  issn = {0370-1573},
  doi = {https://doi.org/10.1016/j.physrep.2023.09.006},
  url = {https://www.sciencedirect.com/science/article/pii/S0370157323003046},
  author = {M. Onorato and Y.V. Lvov and G. Dematteis and S. Chibbaro}
}

@article{onorato2015route,
  author = {Miguel Onorato  and Lara Vozella  and Davide Proment  and Yuri V. Lvov },
  title = {Route to thermalization in the <i>\&\#x3b1;</i>-Fermi\&\#x2013;Pasta\&\#x2013;Ulam system},
  journal = {Proc. Natl. Acad. Sci.},
  volume = {112},
  number = {14},
  pages = {4208-4213},
  year = {2015},
  doi = {10.1073/pnas.1404397112},
  URL = {https://www.pnas.org/doi/abs/10.1073/pnas.1404397112}
}

@article{majda1997one,
  title={A one-dimensional model for dispersive wave turbulence},
  author={Majda, Andrew J and McLaughlin, David W and Tabak, EG1431687},
  journal={J. Nonlinear Sci.},
  volume={7},
  pages={9--44},
  year={1997},
  publisher={Springer},
  URL = {https://doi.org/10.1007/BF02679124}
}

@book{kevrekidis2009discrete,
  title={The discrete nonlinear Schr{\"o}dinger equation},
  author={Kevrekidis, Panayotis G},
  year={2009},
  publisher={Springer Berlin Heidelberg}
}

@book{fibich2015nonlinear,
  title={The nonlinear Schr{\"o}dinger equation},
  author={Fibich, Gadi},
  year={2015},
  publisher={Springer}
}

@article{liu2015renormalized,
  title = {Renormalized phonons in nonlinear lattices: A variational approach},
  author = {Liu, Junjie and Liu, Sha and Li, Nianbei and Li, Baowen and Wu, Changqin},
  journal = {Phys. Rev. E},
  volume = {91},
  issue = {4},
  pages = {042910},
  numpages = {9},
  year = {2015},
  month = {Apr},
  publisher = {American Physical Society},
  doi = {10.1103/PhysRevE.91.042910},
  url = {https://link.aps.org/doi/10.1103/PhysRevE.91.042910}
}

@article{liu2016variational,
  title={Variational approach to renormalized phonon in momentum-nonconserving nonlinear lattices},
  author={Liu, Junjie and Li, Baowen and Wu, Changqin},
  journal={Europhys. Lett.},
  volume={114},
  number={4},
  pages={40002},
  year={2016},
  publisher={IOP Publishing},
  doi = {10.1209/0295-5075/114/40002},
  url = {https://doi.org/10.1209/0295-5075/114/40002}
}

@article{connaughton2005condensation,
  title = {Condensation of Classical Nonlinear Waves},
  author = {Connaughton, Colm and Josserand, Christophe and Picozzi, Antonio and Pomeau, Yves and Rica, Sergio},
  journal = {Phys. Rev. Lett.},
  volume = {95},
  issue = {26},
  pages = {263901},
  numpages = {4},
  year = {2005},
  month = {Dec},
  publisher = {American Physical Society},
  doi = {10.1103/PhysRevLett.95.263901},
  url = {https://link.aps.org/doi/10.1103/PhysRevLett.95.263901}
}

@article{sun2012observation,
  title={Observation of the kinetic condensation of classical waves},
  author={Sun, Can and Jia, Shu and Barsi, Christopher and Rica, Sergio and Picozzi, Antonio and Fleischer, Jason W},
  journal={Nat. Phys.},
  volume={8},
  number={6},
  pages={470--474},
  year={2012},
  publisher={Nature Publishing Group UK London},
  doi = {https://doi.org/10.1038/nphys2278},
  url = {https://doi.org/10.1038/nphys2278}
}

@article{fusaro2019dramatic,
  title = {Dramatic Acceleration of Wave Condensation Mediated by Disorder in Multimode Fibers},
  author = {Fusaro, Adrien and Garnier, Josselin and Krupa, Katarzyna and Millot, Guy and Picozzi, Antonio},
  journal = {Phys. Rev. Lett.},
  volume = {122},
  issue = {12},
  pages = {123902},
  numpages = {7},
  year = {2019},
  month = {Mar},
  publisher = {American Physical Society},
  doi = {10.1103/PhysRevLett.122.123902},
  url = {https://link.aps.org/doi/10.1103/PhysRevLett.122.123902}
}

@article{baudin2023observation,
  title = {Observation of Light Thermalization to Negative-Temperature Rayleigh-Jeans Equilibrium States in Multimode Optical Fibers},
  author = {Baudin, K. and Garnier, J. and Fusaro, A. and Berti, N. and Michel, C. and Krupa, K. and Millot, G. and Picozzi, A.},
  journal = {Phys. Rev. Lett.},
  volume = {130},
  issue = {6},
  pages = {063801},
  numpages = {6},
  year = {2023},
  month = {Feb},
  publisher = {American Physical Society},
  doi = {10.1103/PhysRevLett.130.063801},
  url = {https://link.aps.org/doi/10.1103/PhysRevLett.130.063801}
}

@misc{SuppMat,
  title = {Supplemental Material of "The equilibrium distribution function for strongly nonlinear systems"},
  howpublished = {\url{https://doi.org/xx.xxxx/Supplemental}},
  year = {2025}
}

@article{ramos2023theory,
  title={Theory of localization-hindered thermalization in nonlinear multimode photonics},
  author={Ramos, Alba Y and Shi, Cheng and Fern{\'a}ndez-Alc{\'a}zar, Lucas J and Christodoulides, Demetrios N and Kottos, Tsampikos},
  journal={Commun. phys.},
  volume={6},
  number={1},
  pages={189},
  year={2023},
  publisher={Nature Publishing Group UK London},
  url = {https://doi.org/10.1038/s42005-023-01309-7}
}

@article{pourbeyram2022direct,
  title={Direct observations of thermalization to a Rayleigh--Jeans distribution in multimode optical fibres},
  author={Pourbeyram, Hamed and Sidorenko, Pavel and Wu, Fan O and Bender, Nicholas and Wright, Logan and Christodoulides, Demetrios N and Wise, Frank},
  journal={Nat. Phys.},
  volume={18},
  number={6},
  pages={685--690},
  year={2022},
  publisher={Nature Publishing Group UK London},
  url = {https://doi.org/10.1038/s41567-022-01579-y}
}

@article{xu2008nonequilibrium,
  title = {Nonequilibrium Green's function method for phonon-phonon interactions and ballistic-diffusive thermal transport},
  author = {Xu, Yong and Wang, Jian-Sheng and Duan, Wenhui and Gu, Bing-Lin and Li, Baowen},
  journal = {Phys. Rev. B},
  volume = {78},
  issue = {22},
  pages = {224303},
  numpages = {9},
  year = {2008},
  month = {Dec},
  publisher = {American Physical Society},
  doi = {10.1103/PhysRevB.78.224303},
  url = {https://link.aps.org/doi/10.1103/PhysRevB.78.224303}
}

@article{lee2013generation,
  title={Generation of dispersion in nondispersive nonlinear waves in thermal equilibrium},
  author={Lee, Wonjung and Kova{\v{c}}i{\v{c}}, Gregor and Cai, David},
  journal={Proc. Natl. Acad. Sci.},
  volume={110},
  number={9},
  pages={3237--3241},
  year={2013},
  publisher={National Academy of Sciences},
  url = {https://doi.org/10.1073/pnas.1215325110}
}

@article{tikan2022nonlinear,
  title={Nonlinear dispersion relation in integrable turbulence},
  author={Tikan, Alexey and Bonnefoy, F{\'e}licien and Ducrozet, Guillaume and Prabhudesai, Gaurav and Michel, Guillaume and Cazaubiel, Annette and Falcon, {\'E}ric and Copie, Francois and Randoux, St{\'e}phane and Suret, Pierre},
  journal={Sci. Rep.},
  volume={12},
  number={1},
  pages={10386},
  year={2022},
  publisher={Nature Publishing Group UK London},
  url = {https://doi.org/10.1038/s41598-022-14209-7}
}

@article{lee2019reduced,
  title={Reduced one-dimensional models for wave turbulence system},
  author={Lee, Wonjung},
  journal={J. Nonlinear. Sci.},
  volume={29},
  number={5},
  pages={1865--1889},
  year={2019},
  publisher={Springer},
  url = {https://doi.org/10.1007/s00332-019-09532-9}
}

@article{leisman2022improved,
  title = {Improved effective linearization of nonlinear Schr\"odinger waves by increasing nonlinearity},
  author = {Leisman, Katelyn Plaisier and Zhou, Douglas and Banks, J. W. and Kova\ifmmode \check{c}\else \v{c}\fi{}i\ifmmode \check{c}\else \v{c}\fi{}, Gregor and Cai, David},
  journal = {Phys. Rev. Res.},
  volume = {4},
  issue = {1},
  pages = {L012009},
  numpages = {6},
  year = {2022},
  month = {Jan},
  publisher = {American Physical Society},
  doi = {10.1103/PhysRevResearch.4.L012009},
  url = {https://link.aps.org/doi/10.1103/PhysRevResearch.4.L012009}
}

@article{zhong2023universality,
  title={Universality of light thermalization in multimoded nonlinear optical systems},
  author={Zhong, Qi and Wu, Fan O and Hassan, Absar U and El-Ganainy, Ramy and Christodoulides, Demetrios N},
  journal={Nat. Commun.},
  volume={14},
  number={1},
  pages={370},
  year={2023},
  publisher={Nature Publishing Group UK London},
  url = {https://doi.org/10.1038/s41467-023-35891-9}
}

@article{vsindik2024sound,
  title = {Sound, Superfluidity, and Layer Compressibility in a Ring Dipolar Supersolid},
  author = {\ifmmode \check{S}\else \v{S}\fi{}indik, Marija and Zawi\ifmmode \acute{s}\else \'{s}\fi{}lak, Tomasz and Recati, Alessio and Stringari, Sandro},
  journal = {Phys. Rev. Lett.},
  volume = {132},
  issue = {14},
  pages = {146001},
  numpages = {7},
  year = {2024},
  month = {Apr},
  publisher = {American Physical Society},
  doi = {10.1103/PhysRevLett.132.146001},
  url = {https://link.aps.org/doi/10.1103/PhysRevLett.132.146001}
}

@article{wachtler2016quantum,
  title = {Quantum filaments in dipolar Bose-Einstein condensates},
  author = {W\"achtler, F. and Santos, L.},
  journal = {Phys. Rev. A},
  volume = {93},
  issue = {6},
  pages = {061603},
  numpages = {5},
  year = {2016},
  month = {Jun},
  publisher = {American Physical Society},
  doi = {10.1103/PhysRevA.93.061603},
  url = {https://link.aps.org/doi/10.1103/PhysRevA.93.061603}
}

@article{christodoulides2003discretizing,
  title={Discretizing light behaviour in linear and nonlinear waveguide lattices},
  author={Christodoulides, Demetrios N and Lederer, Falk and Silberberg, Yaron},
  journal={Nature},
  volume={424},
  number={6950},
  pages={817--823},
  year={2003},
  publisher={Nature Publishing Group UK London},
  url = {https://doi.org/10.1038/nature01936}
}

@article{Wang2024Thermalization2,
  title = {Thermalization of Two- and Three-Dimensional Classical Lattices},
  author = {Wang, Zhen and Fu, Weicheng and Zhang, Yong and Zhao, Hong},
  journal = {Phys. Rev. Lett.},
  volume = {132},
  issue = {21},
  pages = {217102},
  numpages = {6},
  year = {2024},
  month = {May},
  publisher = {American Physical Society},
  doi = {10.1103/PhysRevLett.132.217102},
  url = {https://link.aps.org/doi/10.1103/PhysRevLett.132.217102}
}

@article{frisch1953equipartition,
  title = {An Equipartition Principle of Generalized Canonical Ensembles},
  author = {Frisch, Harry L.},
  journal = {Phys. Rev.},
  volume = {91},
  issue = {4},
  pages = {791--793},
  numpages = {0},
  year = {1953},
  month = {Aug},
  publisher = {American Physical Society},
  doi = {10.1103/PhysRev.91.791},
  url = {https://link.aps.org/doi/10.1103/PhysRev.91.791}
}

@book{Pathria2016Statistical,
  title={Statistical mechanics},
  author={Pathria, R. K.},
  year={1996},
  publisher={Butterworth-Heinemann},
}

\end{document}